\documentclass[11pt]{article}

\usepackage[preprint]{acl}

\usepackage{times}
\usepackage{latexsym}

\usepackage[T1]{fontenc}

\usepackage[utf8]{inputenc}

\usepackage{microtype}

\usepackage{inconsolata}

\usepackage{graphicx}
\usepackage{booktabs}
\usepackage{multirow}
\usepackage{tcolorbox}
\usepackage{amsmath} 
\usepackage{enumitem}
\usepackage[normalem]{ulem}
\usepackage{xspace}
\usepackage{xurl}
\usepackage{makecell}

\newcommand{\typeone}{Benchmark Metadata Leakage\xspace}
\newcommand{\typetwo}{Question-Context Leakage\xspace}
\newcommand{\typethree}{Explicit Answer Leakage\xspace}
\newcommand{\stypeone}{BML\xspace}
\newcommand{\stypetwo}{QCL\xspace}
\newcommand{\stypethree}{EAL\xspace}

\usepackage{CJKutf8}

\title{Search-Time Contamination in Deep Research Agents: Measuring Performance Inflation in Public Benchmark Evaluation}

\author{
 \textbf{Yongjie Wang\textsuperscript{1}},
 \textbf{Xinyue Zhang\textsuperscript{3}},
 \textbf{Kunhong Yao\textsuperscript{3}},
 \textbf{Zhiwei Zeng\textsuperscript{1}}, \\
 \textbf{Kaisong Song\textsuperscript{2}}, 
 \textbf{Jun Lin\textsuperscript{2}},
 \textbf{Zhiqi Shen\textsuperscript{3}},
\\
 \textsuperscript{1}Alibaba-NTU Global e-Sustainability CorpLab (ANGEL)\\
 \textsuperscript{2}Tongyi Lab, Alibaba Group, 
 \textsuperscript{3}College of Computing \& Data Science, \\
\texttt{\{yongjie.wang, zhiwei.zeng, zhiqi.shen\}@ntu.edu.sg,\{xinyue020, kunhong002\}@e.ntu.edu.sg} \\
\texttt{\{kaisong.sks,linjun.lj\}@alibaba-inc.com} \\
}

\begin{document}
\maketitle

\begin{abstract}

Public benchmarks enable fair and reproducible evaluation of LLM reasoning, but they become fragile for deep research agents that actively search the web during inference. Such agents may retrieve public benchmark metadata, question context, or even ground-truth answers via web search. This gives rise to Search-Time Contamination (STC), where external retrieval bypasses intended reasoning and inflates measured performance. We systematically study STC in deep research agent evaluation. We define three contamination types with increasing severity, namely \typeone, \typetwo, and \typethree, and develop detection algorithms to identify them and quantify their impact on agent performance. Evaluating modern deep research agents on six public benchmarks, we find that STC is widespread and can inflate performance by up to 4\%. Our findings show that existing evaluations may overestimate true reasoning ability. We therefore advocate contamination-aware practices, including isolated sandboxes, transparent search trajectories, and controlled benchmark access.

\end{abstract}

\section{Introduction}

Evaluating the generative quality and reasoning abilities of large language models (LLMs) \cite{yang2025qwen3,guo2025deepseek,singh2025openai} has largely relied on curated public benchmarks, such as GPQA \cite{rein2024gpqa}, Humanity's Last Exam (HLE) \cite{phan2025humanity}, and BrowseComp \cite{wei2025browsecomp}. Such public benchmarks are essential for conventional evaluation, enabling different models and methods to be assessed against shared reference datasets while supporting transparency, reproducibility, and cumulative progress. 

Sparked by the rapid advances of LLMs , information retrieval is transitioning from passive, keyword-based search to autonomous, agent-driven exploration \cite{nakano2021webgpt, zhang2025deep,wu2025webwalker,jin2025searchr,deng2023mindweb,gou2026mindweb,team2025tongyi}. Unlike traditional search engines that mainly surface static links, deep research agents actively navigate web environments, decompose complex queries into intermediate search tasks through iterative reasoning paradigms such as ReAct \cite{yao2023react}, and synthesize fragmented evidence into citation-grounded responses. This active retrieval capability substantially improves their performance on challenging benchmarks. For example, on HLE, web-enabled Grok 4 improves from 25.4\% to 38.6\%, while Grok 4 Heavy further reaches 44.4\%, nearly doubling the original score.

Despite these gains, applying existing benchmark evaluation protocols directly to deep research agents raises a critical methodological concern. Since benchmark instances are often curated from web corpora or released through public dataset hosting platforms such as HuggingFace and GitHub, an agent may retrieve the original test question, answer cues, or even the ground-truth answer during inference. This phenomenon is referred to as Search-Time Contamination (STC) \cite{han2025searchtime}: a failure mode where retrieved benchmark artifacts allow the agent to bypass the intended reasoning process, leading to poor generalization on practice questions that are not available on the web. As illustrated in Figure \ref{fig:pipeline}, once Tongyi Deep Research retrieves the ground-truth label from a Chegg webpage, it deviates from its initial reasoning path and directly adopts the leaked answer. Such cases lead to apparently correct responses that do not faithfully reflect the agent's underlying reasoning capability.

Systematically quantifying STC-induced performance inflation is therefore essential for establishing fair, trustworthy, and rigorous evaluation protocols for agentic systems. In this paper, we analyze the mechanisms through which STC occurs and classify it into three fine-grained types along a spectrum of severity: (1) \textit{\typeone}, where the agent's search behavior is directed toward locating benchmark artifacts rather than solving the task; (2) \textit{\typetwo}, where retrieved content reveals the surrounding context or source of the benchmark instance without explicitly exposing the answer; and (3) \textit{\typethree}, where the ground-truth label or answer is directly retrieved during inference. Correspondingly, we introduce detection algorithms to identify these contamination types and quantify their impacts on measured agent performance from both question level and turn levels. 

We mainly investigated STC on clinical and medical benchmarks, where questions often require multi-hop reasoning, deductive inference, and precise factual recall. Because many public medical benchmarks are widely reproduced across repositories, examination platforms, question banks, educational websites, and forums, they provide a high-stakes and highly exposed setting for studying STC. Based on our study, all evaluated benchmark datasets exhibit varying degrees of STC. The issue is particularly severe for early benchmark MedMCQA, where nearly one quarter of the questions can have their answers retrieved from the web. STC-induced performance inflation can reach up to 4\% on HLE biological and chemical subsets. Once the leaked answer is retrieved, the agent achieves both higher accuracy and faster convergence, reaching over 95\% accuracy regardless of medical task difficulty and rapidly shifting from its previous reasoning trajectory to the leaked answer. Similar leakage patterns are also observed in commercial deep research systems such as Gemini Deep Research: the agent visits webpages that explicitly contain both the original question and its corresponding answer for 60\% of the 100 sampled MedQA questions. 

Our findings indicate that the improved performance observed in deep research agents may be partially attributable to web-based answer leakage, rather than entirely reflecting genuine reasoning improvement. We therefore advocate contamination-aware evaluation practices for deep research agents, including isolated sandboxes, which prevent agents from retrieving benchmark artifacts or answer keys during inference; transparent search trajectories, which allow auditors to inspect search queries, retrieved URLs, visited pages, and summarized evidence for potential contamination; and controlled benchmark access, such as private test sets or dynamically generated instances, further reduces the risk that public web exposure turns evaluation into retrieval-based answer matching rather than genuine reasoning.

\section{Preliminary}
\textbf{Deep Research} refers to an autonomous agentic system that combines step-by-step reasoning with external actions, such as web search, to solve complex decision-making tasks. The agent executes a dynamic, iterative trajectory, usually including the following key components:
\begin{itemize}[topsep=1pt,leftmargin=10pt,itemsep=0pt]
\item \textbf{Reasoning ($\tau_t$)}: The agent accumulates the execution history and reasons over it to deduce the next step action or generate the final answer.
\item \textbf{Action ($a_t$)}: The agent executes an external operation to interact with the environment, such as submitting a query to a web search engine.
\item \textbf{Observation ($o_t$)}: The agent receives feedback from the environment after executing an action and uses it to update its internal context for subsequent reasoning.
\end{itemize}


Following the ReAct framework \cite{yao2023react}, we represent an agent trajectory as a sequence of reasoning, action, and observation steps:
\begin{align}
    \mathcal{H}_T = \{q, (\tau_0, a_0, o_0), ..., (\tau_T, a_T)\}
\end{align}
At the terminal step $T$, the agent provides the final answer $a_T$. For each intermediate step $t$, the agent learns to optimize the policy $\pi(\cdot| \mathcal{H}_{t-1})$, aiming to produce the optimal thought $\tau_t$ and action $a_t$:
\begin{align}
    (\tau_t, a_t) \sim \pi(\cdot| \mathcal{H}_{t-1})
\end{align}

Web Search is one widely used external tool for gathering broad knowledge and synthesizing information \cite{gou2026mindweb,nakano2021webgpt,jin2025searchr,team2025tongyi}. It provides access to more up-to-date, diverse, and open-domain information, making it particularly useful for complex tasks that require recent facts, cross-source verification, or evidence aggregation. Typically, web search consists of two atomic operations: 
(i) \textbf{Search}, submitting a query to a search engine and receives a ranked list of relevant URLs together with short snippets; and 
(ii) \textbf{Visit} (\textit{Browse}), opening selected webpages from the search results, reads their content, and extracts or summarizes task-relevant evidence. This search-and-visit mechanism allows deep research agents to collect complementary evidence from multiple sources and synthesize them into a final response.

\textbf{Search-Time Contamination (STC)} was first formally discussed in~\cite{han2025searchtime}. It refers to a failure mode in which the retrieval step of a deep research agent returns answer-revealing web content derived from, or closely overlapping with, the evaluation set itself. Such answer-revealing evidence allows the agent to bypass the intended reasoning process through retrieval shortcuts, leading to poor generalization on unseen questions that are not available on the web. Measuring STC is therefore essential for separating genuine reasoning capability from contamination-induced performance inflation and for ensuring trustworthy evaluation.


Prior work~\cite{han2025searchtime} detects search-time leakage through URL matching against Hugging Face. While this offers an important initial step, it mainly captures coarse-grained benchmark exposure and leaves other contamination forms under-examined. First, repository-level matching may miss leakage from broader web sources, such as forums, educational websites, question banks, or informal discussions. Second, a repository hit does not necessarily imply answer leakage, as benchmark queries and labels may be separated, hidden, or hosted in different locations. Therefore, repository matching alone is insufficient to distinguish different STC mechanisms, which may range from retrieving benchmark metadata and question context to directly retrieving gold-standard answers. 

\section{Methodology}
In this section, we formalize Search-Time Contamination (STC) with a three-level taxonomy: \typeone, \typetwo, and \typethree. We then design specific detection mechanisms to identify each type of contamination.

\begin{figure*}
    \centering
    \includegraphics[width=\linewidth]{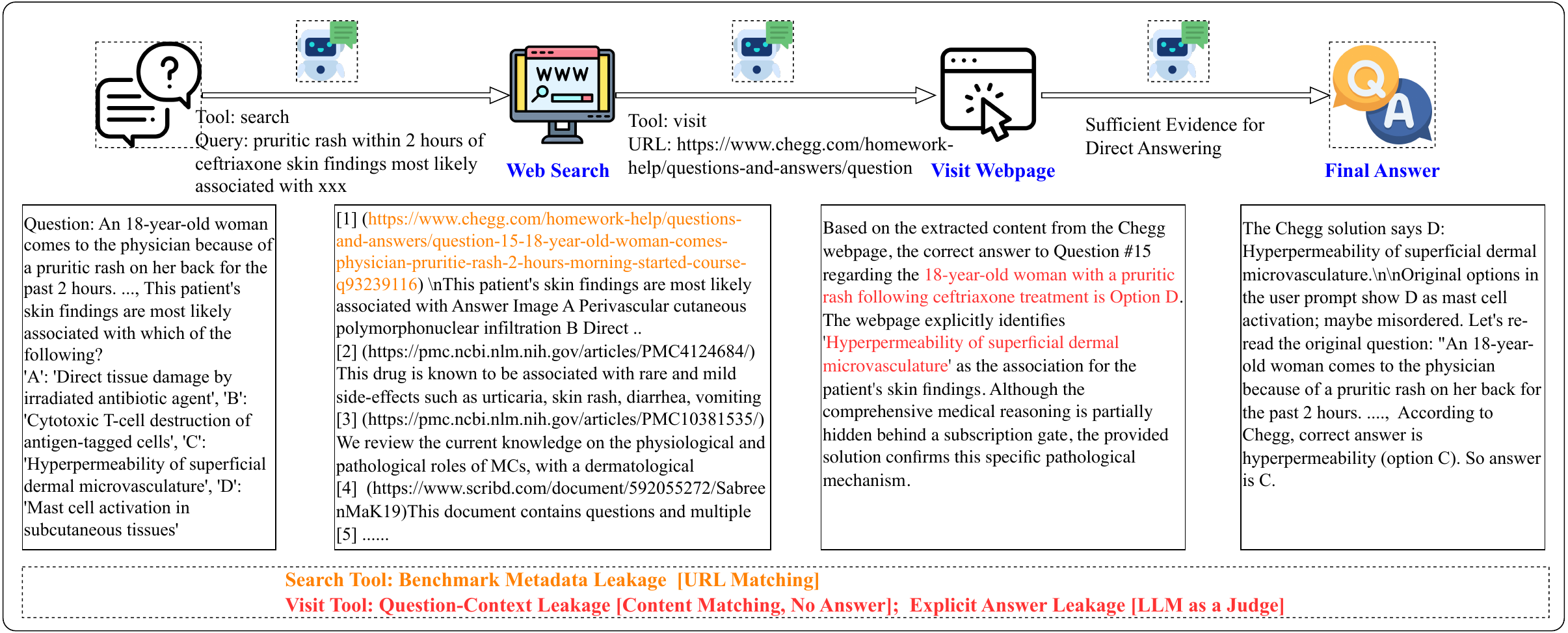}
     \caption{Illustration of clinical search-time contamination with Tongyi Deep Research. After retrieving the answer, the agent directly jumps to prediction, bypassing the intended reasoning process. This example can be found in MedQA output in our released repository, Line 715.}
    \label{fig:pipeline}
    \vspace{-0.5cm}
\end{figure*}

\subsection{STC Taxonomy}
During iterative search, the agent may encounter benchmark-related artifacts at different stages, including top-ranked search results and subsequently visited webpages. As such, we define three types of STC according to the agent operation involved and the severity of the exposed information.

\begin{itemize}[topsep=1pt,leftmargin=10pt,itemsep=0pt]
    \item \textbf{\typeone (\stypeone)} occurs when the search step returns URLs that expose benchmark-specific metadata, such as benchmark names or question identifiers, rather than useful knowledge sources. Although the webpage content has not yet been visited at this stage, the retrieved URLs already suggest that the agent's search process has been steered away from task-relevant knowledge gathering toward benchmark-related artifacts.
    \item \textbf{\typetwo (\stypetwo)} occurs when the retrieval process returns documents containing the exact phrasing or highly specific context of the test question, but not its ground-truth label. For example, the agent may retrieve the original article or source passage from which a QA pair was constructed. Such context can provide external priors and substantially narrow the search space, thereby allowing agents to shortcut the intended independent reasoning process. 
    \item \textbf{\typethree (\stypethree)} occurs when the retrieval process returns content that contain both the evaluation query $x$ and its corresponding ground-truth label $y$. In this scenario, the reasoning chain completely collapses into a trivial key-value extraction task, rendering the benchmark evaluation fundamentally invalid.
\end{itemize}
Since we define contamination types at the step level, a single question-answering trajectory may exhibit multiple STC types, as shown in Figure \ref{fig:pipeline}. 

\subsection{Detection Algorithm}
\textbf{\stypeone Detection.} To infer an LLM’s search intent from returned URLs, we employ a regular-expression-based URL matching method. For each query phrase, the search engine returns top relevant websites and snippets. We detect whether the returned URLs are suspicious by applying regular-expression matching over two groups of websites: (1) common source websites, including data-hosting platforms (e.g. Hugging Face and GitHub), which often provide free access to open-source datasets; and medical examination URLs (e.g. Quizlet), which serve as original data sources for benchmarks and expose exam questions or answers directly; and (2) benchmark-specific websites, which contain dataset names, and known dataset source patterns. The complete set of matching templates is provided in Appendix \ref{sec:urls}. The occurrence of any of these two groups of search results serves as an indicator that the LLM seeks dataset-level metadata, rather than question-relevant medical knowledge.

\noindent \textbf{\stypetwo Detection.}
We detect question-context leakage by measuring the lexical overlap between the question and the retrieved content. Specifically, we compute the longest common substring between the retrieved content and the question, and normalize it by the question length to account for length variations. A higher normalized overlap indicates that the question itself is more likely to have appeared in the retrieved content. However, string matching alone cannot determine whether the retrieved content also exposes the corresponding answer. Therefore, we combine EAL detection to exclude cases involving explicit answer leakage. 

\noindent \textbf{\stypethree Detection.} To further identify whether the retrieved content explicitly contains the answer, we adopt LLM-as-a-Judge~\cite{zheng2023judging}. 
We prompt DeepSeek V4 Pro \cite{deepseekai2026deepseekv4} to determine whether the retrieved content provides direct evidence for the answer.  The prompt is provided in Figure~\ref{fig:prompt_type3} of the Appendix section.

We evaluate the correctness of automatic detection by comparing it against human-annotated ground-truth labels on the Medbullets5op and MedQA datasets. 
For Medbullets-5op, we compare the EAL detection results with finalized human annotations and find that the detector achieves 83.3\% recall and 100\% precision. 
For MedQA, due to the larger dataset size, we conduct human post-checking only on the detected EAL cases and therefore report precision only; the detection algorithm achieves 94.85\% precision. The detailed results are reported in Appendix~\ref{apd:human_automatic}.

\section{Evaluation}
 To systematically assess this complex vulnerability and its impact on final prediction performance, we introduce an evaluation framework from two complementary perspectives: a \textbf{question-level evaluation} and a \textbf{turn-level evaluation}.

\subsection{Question-level Evaluation} Since a single question may be associated with multiple STC types, our main strategy is to compare agent performance across subsets split by specific STC types. We assume that questions in the compared subsets have comparable difficulty, so the observed performance gap can be primarily attributed to the presence of STC events. 

To assess the impact of BML, we partition the dataset into BML-triggered and non-BML subsets and compare agent performance between them. Since QCL and EAL occur at the visit stage, they are inherently conditioned on BML, where a relevant URL has already been retrieved. We therefore further split the BML-triggered subset by the presence or absence of QCL and EAL, yielding four subgroups, and report agent performance for each. 

\subsection{Turn-level Evaluation} 
The turn-level evaluation investigates how leakage triggered during a search step impacts the final prediction and analyzes the interaction among the three STC types. 

\textbf{Prediction Shift.} Because agents may generate tentative predictions during the search process, we analyze how their answers evolve when an STC occurs. Specifically, assuming a contamination event is triggered at turn $t$, we quantify the \textbf{prediction shift} by comparing the agent's prediction at turn $t$ against its prediction at turn $t+1$. Notably, if the LLM fails to output any prediction at a given turn, it is strictly evaluated as incorrect. 

\textbf{Dynamic Impact on Predictions.} We treat the agent's step-by-step search process as a survival problem where each agent turn corresponds to a time step, the event of interest is that the agent reaches a correct prediction, and STC events are modeled as time-dependent covariates, indicating whether a specific contamination type has occurred by a given turn. This formulation allows us to model not only whether the agent eventually answers correctly, but also when the correct prediction emerges after an STC event is triggered, using survival analysis models in public health.

To estimate how STC affects this process, we use a time-varying \textit{Cox proportional hazards model} \cite{cox1972regression}, referring to Section \ref{apd:cox} in the Appendix. The estimated hazard ratio measures how the occurrence of an STC event changes the instantaneous likelihood of reaching a correct prediction at that turn. A hazard ratio larger than 1 indicates that the agent is more likely to reach the correct answer after the corresponding STC event. \textit{For example, a remarkably high HR for \stypethree contamination indicates that the agent outputs the correct prediction quickly after answers are explicitly leaked. Conversely, a lower HR implies a delayed or weaker correlation between the contamination event and the successful prediction.}

\textbf{Cascading Effect of STC Types.}  In the search trajectory, contamination events are rarely isolated; the occurrence of one leakage type may catalyze another. We therefore investigate the internal sequential dynamics among these STC states. We use Kaplan--Meier curves to analyze STC escalation after \stypeone occurs. Specifically, we reset the time origin to the first \stypeone turn and measure the number of subsequent turns until the first occurrence of \stypetwo or \stypethree. Furthermore, to evaluate the forward cascading effect from question to answer, we set \stypethree as the target event and \stypetwo as the covariate. This configuration quantifies whether encountering the exact question early in the search catalyzes the subsequent retrieval of the explicit answer. 

\begin{table*}[ht]
    \centering 
    \caption{Turn-level and question-level STC statistics across benchmarks.}
    \label{tab:stc_statistics}
    \resizebox{0.9\linewidth}{!}{
    \begin{tabular}{l|ccccc|ccccc} \toprule
    \multirow{2}{*}{Dataset} & \multicolumn{5}{c|}{Turn-level Statistics} & \multicolumn{5}{c}{Question-level Statistics}  \\ \cline{2-11}
     & Avg Turn & No STC & \stypeone & \stypetwo & \stypethree & No STC & \stypeone & BML\&\stypetwo & BML\&\stypethree & All Types \\ \midrule  
     MedQA  & 6.55 & 7427 & 834 & 44 & 39 & 872 & 351 & 16 & 23 & 11 \\
     MMLU   & 6.23 & 2873 & 431 & 27 & 56 & 382 & 114 & 6 & 32 & 10 \\
     MedMCQA & 15.87 & 24274 & 7543 & 569 & 769 & 533 & 934 & 115 & 324 & 183\\ 
     MedXpertQA & 10.22 & 21973 & 3031 & 1 & 25 & 1456 & 971 & 0 & 22 & 1\\ 
     HLE-149 & 14.23 & 1917 & 192 & 6 & 5 & 77 & 66 & 3 & 1 & 2\\ 
     Medbullets5op & 9.85 & 2612 & 296 & 18 & 10 & 203 & 81 & 5 & 2 & 7\\     \bottomrule
    \end{tabular}} 
    \vspace{0.1cm}

\centering
\caption{Main results of Tongyi-DeepResearch. Tongyi-DeepResearch* disables web search; results are reported on the full dataset, subsets splited by \stypeone occurrence, and by \stypetwo/\stypethree co-occurrence within the \stypeone subset.}
\label{tab:main_results}
\resizebox{0.9\textwidth}{!}{
\begin{tabular}{lcccccc}
\toprule
{Name} &  MedQA & MedMCQA & MMLU  & MedXpertQA & HLE-149 & Medbullets5op\\
\midrule
\multicolumn{7}{c}{Full Dataset Evaluation} \\  \midrule 
Qwen3-30B-A3B & 83.58\% & 70.27\% & 89.72\%  & 21.80\% & 12.75\% & 67.45\% \\
Tongyi-DeepResearch* & 89.00\% &72.47\% & 91.96\%  & 28.45\% & 15.44\% & 73.15\%\\
Tongyi-DeepResearch & 91.28\% & 87.34\% & 94.86\%  & 40.61\% & 24.83\% & 76.17\% \\
\midrule 
\multicolumn{7}{c}{Split by \typeone (\stypeone) Occurrence} \\  \midrule
Without \stypeone & 92.78\% & 90.24\% & 94.24\%  & 37.64\% & 20.78\% & 74.38\% \\
With \stypeone & 88.03\% & 86.12\% & 95.68\%  & 44.97\% & 29.17\% & 80.00\% \\
\midrule 
\multicolumn{7}{c}{Breakdown within the \stypeone Group} \\  \midrule
With \stypeone only & 88.03\% & 84.37\% & 94.74\% & 44.28\% & 27.27\% & 80.25\% \\
With \stypeone and \stypetwo & 68.75\% & 61.74\% & 100\% & N/A & 0\% & 60\% \\
With \stypeone and \stypethree & 95.65\% & 97.53\% & 96.88\% & 72.73\% & 100\% & 50\% \\
With \stypeone, \stypetwo and \stypethree & 100\% & 90.16\% & 100\% & 100\% & 100\% & 100\% \\
\bottomrule
\end{tabular}}
\end{table*}

\section{Experiment}
\textbf{Datasets.} We evaluate clinical search-time contamination on six medical question-answering benchmarks: MedQA \cite{jin-2021-medqa}, MMLU medical subsets \cite{hendrycks2021measuring}, MedMCQA \cite{pal-2022-medmcqa}, MedXpertQA \cite{zuo2025medxpertqa}, HLE medical subsets \cite{phan2025humanity}, and Medbullets5op \cite{chen-2025-benchmarking}. For MMLU and MedMCQA, we exclude the shortest half of questions, as they are more likely to require only factual recall rather than complex reasoning, inspired by \cite{thapa2025disentangling}. More dataset details are provided in Appendix Section \ref{sec:dataset}.

We directly report accuracy for multiple-choice questions. 
For the HLE subset, which includes open-ended questions, we use LLM-as-a-judge to estimate accuracy following the standard HLE evaluation protocol. Note that our goal is not to compare the relative performance of different deep research agents. Instead, we focus on measuring the severity of STC and the extent to which performance changes after rectification. 

\noindent \textbf{Deep Research Agents.} We evaluate a representative set of deep research agents spanning both open-source models and commercial research systems, including \textbf{Tongyi Deep Research} \cite{team2025tongyi}, \textbf{Step Deep Research} \cite{hu2025stepdeepresearchtechnicalreport}, \textbf{Gemini Deep Research} \cite{google2026gemini} and \textbf{Valyu Deep Research} \cite{valyu2026deepresearch}. In addition, we include \textbf{Qwen3-30B-A3B} \cite{qwen3technicalreport}, the base model of Tongyi Deep Research, to assess the underlying model capability before applying the full deep research agent pipeline. More details of agents are provided in Appendix Section \ref{apd:agents}.

Our main experiments are conducted on Tongyi Deep Research because it is cost-effective, publicly accessible, and provides fully observable execution traces, allowing us to inspect search queries, visited webpages, intermediate summaries, and reasoning steps. To examine the generalizability of our findings, we further evaluate several other deep research agents. All experiment results are released at GitHub \footnote{https://anonymous.4open.science/r/Search-Time\_Contamination-25F2/}.

\subsection{Statistics Report}
Table \ref{tab:stc_statistics} reports the turn-level and question-level statistics of STC types. 
We observe that STC appears at different severity levels across evaluated datasets, indicating that STC is not an isolated phenomenon but a common risk in evaluation. 

Earlier benchmarks such as MedQA, MedMCQA and MMLU show stronger contamination signals, likely because they have been available longer and are more widely indexed, mirrored, and discussed online. In particular, medical-exam-derived datasets such as MedQA and MedMCQA are frequently redistributed through exam-preparation websites, forums, repositories, and social platforms, making them more susceptible to retrieval during web search. Notably, benchmark recency does not eliminate STC risk. Even recently released datasets such as MedXpertQA and HLE-149 show leakage signals, mainly due to exposure through LLM-related publications, data hosting platforms, or online forum discussions. 

\subsection{Main Results}

The question-level evaluation results are reported in Table \ref{tab:main_results}. Disabling the web search, the agent consistently improves over the base Qwen3-30B-A3B model, showing that finetuning enhances the model's underlying reasoning capability. Enabling web search further improves performance, but these gains are partially attributable to STC. 

The results on the splits induced by \stypeone do not always indicate performance inflation. Although the \stypeone-triggered subset performs better on four benchmarks, it performs worse on two others, contrasting with prior conclusions based on repository matching \cite{han2025searchtime}. This suggests that URL-level benchmark metadata leakage alone is insufficient to determine whether the agent actually benefits from contamination.

The fine-grained breakdown within the \stypeone group shows that the actual performance impact depends on whether the agent subsequently retrieves explicit answers. 
Subsets involving \stypethree tend to achieve much higher accuracy, such as 100\% on the HLE subsets and 73.91\% on MedXpertQA, substantially exceeding average performance. 
In contrast, \stypetwo alone does not consistently improve performance, suggesting that query-context exposure may introduce partial or noisy priors rather than direct answer shortcuts. These results demonstrate the necessity of a multi-level STC taxonomy and highlight the need for greater awareness of STC risks in search-enabled agent evaluation. 

\subsection{Turn-Level Prediction Dynamics}
Table~\ref{tab:turn_level} reports the prediction dynamics. We observe a sharp improvement in prediction accuracy after the agent retrieves the answer. Interestingly, answer options on external websites may appear in shuffled orders, which can initially mislead the LLM. After one or two additional reasoning turns, however, the agent may correct its prediction, as shown in Figure~\ref{fig:pipeline}. This explains why performance after EAL is lower than the final prediction performance (80\% vs 88.7\% on Medbullet5op datasets).  In contrast, BML and QCL exhibit more varied behaviors. Moreover, when the LLM fails to output a prediction at an intermediate turn, we strictly count that turn as incorrect, which makes intermediate prediction dynamics noisier. Providing only web snippets or question-related context without directly exposing the answer may not consistently improve prediction accuracy. 
\subsection{Impact on Predictions}
Table \ref{tab:type_importance} presents the time-varying Cox regression results. 
A clear pattern is that \stypethree has the strongest and most consistent positive association with correct prediction, indicating that explicit answer leakage substantially increases the likelihood that the agent reaches a correct prediction in subsequent search turns. Its hazard ratios are consistently larger than 1, ranging from 2.20 on Medbullets5op to 8.92 on HLE-149, and are statistically significant on most datasets. By contrast, \stypeone does not consistently improve prediction. 
Its hazard ratios are below 1 on $5$ benchmarks, suggesting that benchmark metadata leakage alone is not sufficient to help the agent answer correctly. \stypetwo shows mixed effects: it is negatively associated with correct prediction on several benchmarks and is not statistically significant on others. 

\subsection{Cascading Effect between STC Types}
\begin{figure*}[t]
    \centering
    \includegraphics[width=0.45\linewidth]{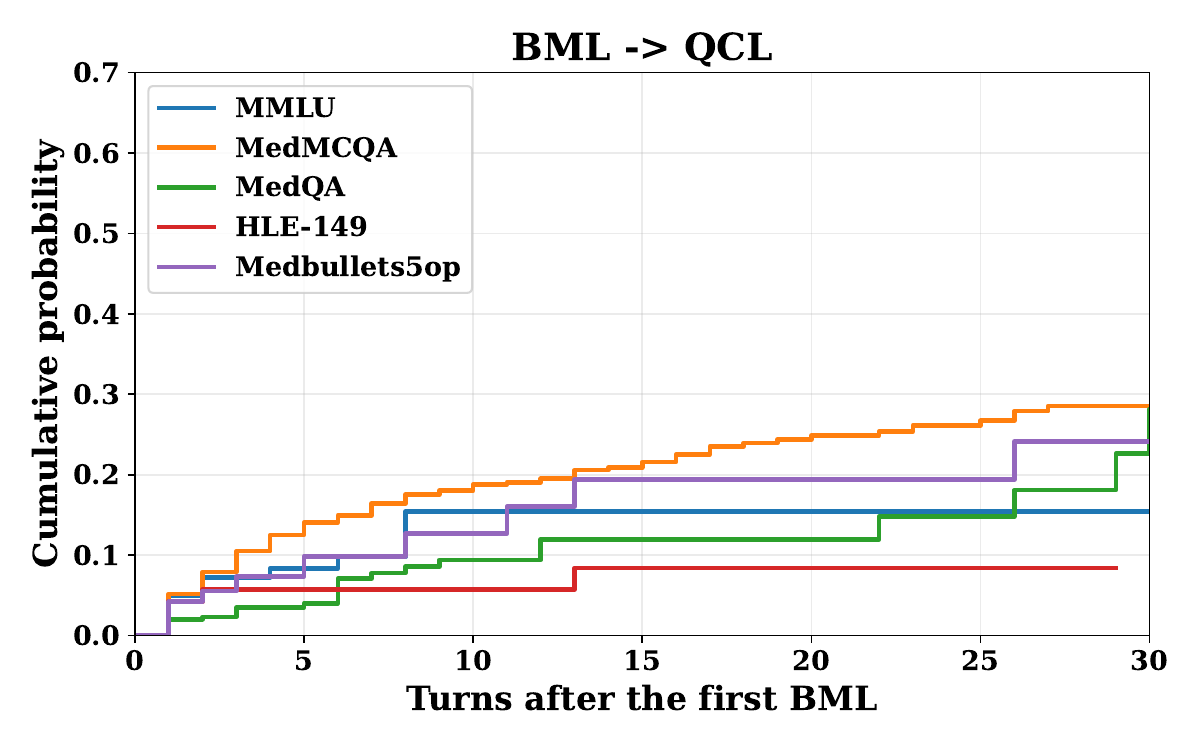}
    \includegraphics[width=0.45\linewidth]{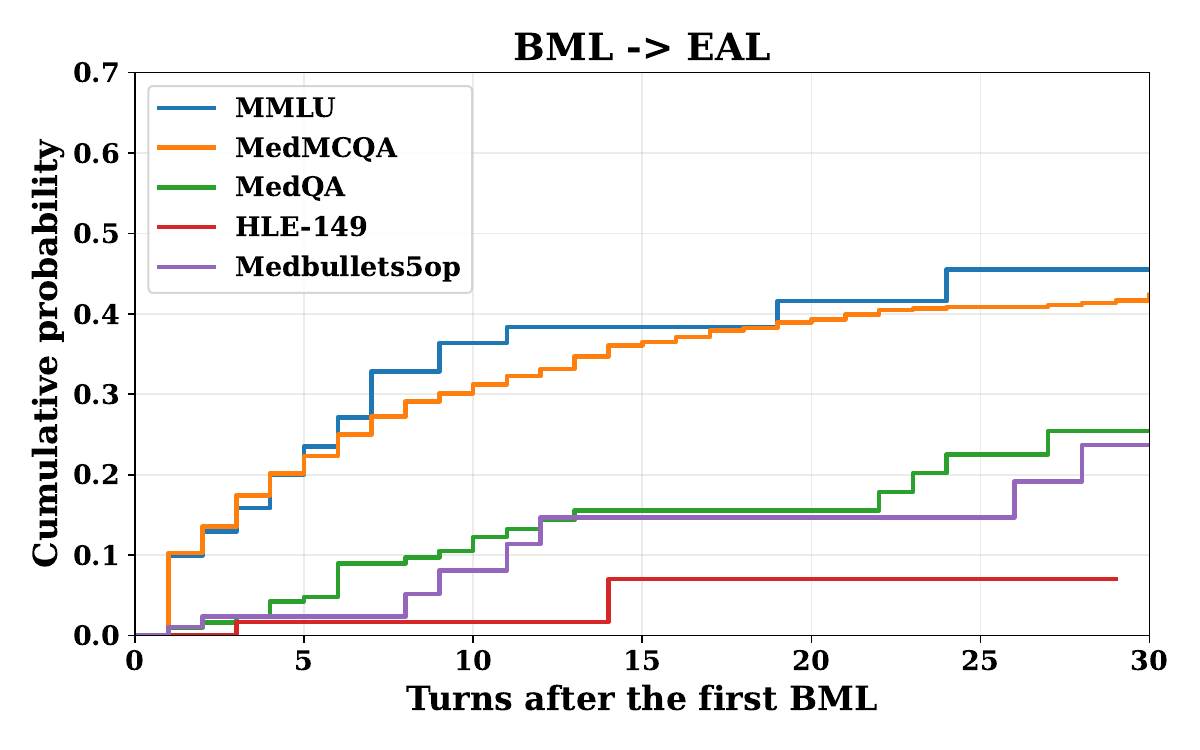}
    \caption{Kaplan--Meier curves of STC escalation after \stypeone.}
    \label{fig:km}
\end{figure*}

Figure~\ref{fig:km} shows the Kaplan--Meier curves of STC escalation after the first BML event. Across datasets, BML is more likely to escalate into EAL than into QCL, suggesting that weak contamination observed in search snippets can serve as an early warning signal for subsequent explicit answer leakage during webpage visits. This pattern is especially pronounced on MMLU and MedMCQA, while HLE-149 shows a lower escalation probability. Table~\ref{tab:type_interaction} shows that \stypetwo significantly increases the hazard of subsequent \stypethree across most datasets, with hazard ratios ranging from 2.50 to 6.74. This suggests that once \stypetwo occurs, the agent is more likely to escalate to \stypethree in later turns. The only exception is HLE-149, where the confidence interval is wide and the effect is not statistically significant, likely due to limited event counts.

\begin{table*}[ht]
\caption{Performance comparison between the current turns and after detected STC events.}
\resizebox{0.9\textwidth}{!}{
\label{tab:turn_level}
\begin{tabular}{l|ccc|ccc|ccc}
\toprule
\multirow{2}{*}{Dataset} &  \multicolumn{3}{c}{\typeone} & \multicolumn{3}{c}{\typetwo} & \multicolumn{3}{c}{\typethree} \\ \cline{2-10} 
 & \#Samples & Before & After & \#Samples & Before & After & \#Samples & Before & After \\
\midrule  
MedQA & 834 & 48.44\% & 30.94\% & 44 & 4.55\% & 22.73\% & 39 & 7.69\% & 89.74\% \\
MMLU & 431 & 52.67\% & 36.89\% & 27 & 18.52\% & 48.15\% & 56 & 17.86\% & 82.14\% \\
MedMCQA & 7543 & 27.71\% & 16.62\% & 569 & 14.24\% & 25.66\% & 769 & 19.25\% & 79.45\% \\
MedXpertQA & 3031 & 7.06\% & 6.20\% & 1 & 0\% & 0\% & 25 & 8\% & 48\%  \\
HLE-149 & 192 & 9.90\% & 7.81\% & 6 & 0\% & 0\% & 5 & 20\% & 100\%  \\
Medbullets5op & 296 & 25.34\% & 16.22\% & 18 & 0\% & 11.11\% & 10 & 0\% & 80\% \\
\bottomrule
\end{tabular}}
 \vspace{0.2cm}
\centering
\caption{
Time-varying Cox regression results for the association between each STC type and correct prediction. 
We report hazard ratios (HRs), 95\% confidence intervals (CIs), and $p$-values.}
\resizebox{0.9\textwidth}{!}{
\label{tab:type_importance}
\begin{tabular}{l|ccc|ccc|ccc}
\toprule
\multirow{2}{*}{Dataset} &  \multicolumn{3}{c|}{\stypeone $\to$ Accuracy} & \multicolumn{3}{c|}{\stypetwo $\to$ Accuracy} & \multicolumn{3}{c}{\stypethree $\to$ Accuracy} \\ \cline{2-10}
& HR & 95\% CI & p-value  & HR & 95\% CI & p-value  & HR & 95\% CI & p-value  \\ 
\midrule
MMLU & 0.45 & [0.36, 0.55] & < 0.005 & 0.78 & [0.47, 1.28] & 0.32 & 2.21 & [1.54, 3.18] &  <0.005\\
MedMCQA & 0.62 & [0.56, 0.69] & <0.005 & 0.72 & [0.63, 0.82] & <0.005 & 2.27 & [2.03, 2.53] & <0.005\\
MedQA & 0.68 & [0.60, 0.78] & <0.005 & 0.64 & [0.43, 0.97] & 0.035 & 4.21 & [2.89, 6.13] & <0.005\\
MedXpertQA & 0.85 & [0.76, 0.96] & 0.0069 & 2.99 & [0.41, 22.05] & 0.28 & 2.49 & [1.53, 4.05] & <0.005\\
HLE-149 & 1.06 & [0.62, 1.84] & 0.82 & 0.94 & [0.29, 3.11] & 0.92 & 8.92 & [2.46, 32.35] & <0.005\\
Medbullets5op & 0.78 & [0.59, 1.02] & 0.07 & 1.10 & [0.57, 2.10] & 0.79 & 2.20 & [0.99, 4.86] & 0.052 \\
\bottomrule
\end{tabular}}
\end{table*}

\begin{table}[tb]
  \centering
\caption{Time-varying Cox regression results between \stypetwo and \stypethree. We report HR, 95\% confidence interval, and $p$-values.}
\resizebox{0.8\linewidth}{!}{
\label{tab:type_interaction}
\begin{tabular}{l|ccc}
\toprule
\multirow{2}{*}{Dataset} &  \multicolumn{3}{c}{\stypetwo $\to$ \stypethree}  \\ \cline{2-4}
   & HR & 95\% CI & p-value  \\ \midrule
MMLU & 3.51 & [1.56, 7.91] &  <0.005 \\
MedMCQA & 2.50 & [2.04, 3.06] & <0.005\\
MedQA & 2.95 & [1.43, 6.09]  & <0.005\\
HLE-149 & 2.12 & [0.35, 12.98] & 0.42 \\
Medbullets5op & 6.74 & [2.46,18.50] & <0.005 \\
\bottomrule
\end{tabular}} 
\vspace{-0.2cm}
\end{table}

\begin{table}[tb]
\centering
\caption{
STC severity across deep research agents on the first 100 MedQA questions.}
\label{tab:general_results}
\resizebox{0.9\linewidth}{!}{
\begin{tabular}{c|cc|c}
\toprule
\multirow{2}{*}{Models} & \multicolumn{2}{c|}{Web Search} & \multirow{2}{*}{\makecell{Leakage\\Rate}} \\
\cline{2-3}
& On & Off & \\
\midrule 
Gemini Deep Research & 99\% & 97\% & 60\% \\
Step Deep Research & 93\% & 92\% & 9\% \\ 
Valyu Deep Research & 87\% & N/A & 0\% \\
\bottomrule
\end{tabular}}
\vspace{-0.2cm}
\end{table}

\subsection{Generalizability of STC}
Here, we verify whether search-time contamination is a ubiquitous vulnerability for other agents. As the internal search trajectories of proprietary agents are opaque, we only have access to their final predictions, summaries, and the list of visited URLs. Therefore, We first match each question against visited webpage content, then manually verify answer exposure. If an visited source contains both exact answer and questions, we label the instance as answer leakage. 

As shown in Table \ref{tab:general_results}, Gemini exhibits an alarmingly high leakage rate of 60\% when web search is enabled. Although its underlying agent model is already strong, such a high leakage rate raises serious concerns. Step Deep Research shows a more moderate level of contamination, with a leakage rate of 9\%. One possible reason is that its search infrastructure is more Chinese-oriented, which may make it less effective at retrieving English benchmark artifacts from the web. Conversely, Valyu shows an exceptional leakage rate of 0\% on MedQA.  This may be due to its unique search infrastructure, which appears to retrieve evidence from restricted or curated sources, such as PubMed, rather than general webpages.  However, this design does not eliminate STC risk: when evaluated on PubMedQA, whose questions are curated from PubMed articles, Valyu also exhibits a 78\% leakage rate (seen in Appendix Section \ref{apd:valyu}).

\section{Discussion}
Our experiments lead to a critical conclusion: the exceptional performance exhibited by deep research agents is, at least partially, an artifact of web-based answer leakage. How to maintain a robust, contamination-free knowledge-gathering process for measuring an agent's reasoning capabilities remains a pivotal challenge.

First, future evaluations should be conducted in an isolated knowledge sandbox, such as ToolUniverse~\cite{gao2025democratizing}, to ensure fair comparisons. All evaluated agents should retrieve information from the same sandbox, differing only in their underlying model or reasoning policy.

Current commercial systems return only a synthesized summary and reference URLs rather than the full search trajectory, which makes STC events difficult to detect. Offering retrieved URLs, search queries, and intermediate evidence would enable users and evaluators to examine whether the final answer is grounded in legitimate reasoning or merely reflects leaked benchmark artifacts. 

Third, benchmark access must be strictly controlled rather than fully indexed by search engines or vulnerable to automated crawlers. To enforce this, dataset repositories should implement mechanisms such as gated access and mandatory user registration. Furthermore, data use agreements must explicitly prohibit the unauthorized redistribution of benchmark instances. 

\section{Conclusion}
This work identifies search-time contamination (STC) as a critical reliability threat in evaluating deep research agents. 
We categorize STC into three types---\typeone, \typetwo, and \typethree---and introduce corresponding detection methods to assess their prevalence and influence in public medical benchmarks. 
Our results show that modern agents are vulnerable to varying degrees of contamination, which can inflate reported performance and even alter model rankings. 
These findings highlight the need for strict knowledge sandboxes to separate genuine reasoning ability from superficial retrieval of benchmark artifacts.

\section*{Limitations}
While this study exposes important vulnerabilities in existing deep research agents, several limitations remain. 

First, our URL matching strategy for BML detection focuses on commonly observed websites that are highly likely to contain benchmark artifacts. However, websites outside these predefined matching patterns may also contain benchmark questions, leading to imperfect recall and a potential underestimation of BML prevalence. 

Second, our experiments are limited to clinical benchmarks, which are reasoning-intensive and widely available on the web. Similar issues have also been observed in general QA tasks~\cite{han2025searchtime}, suggesting that STC is a broader phenomenon. However, more task domains should be examined in future work for rigorous evaluation.

Third, evaluating commercial agents is costly and time-consuming, so we include only three commercial systems. However, search-time contamination mainly comes from open-web search over publicly indexed benchmark artifacts. Without sandboxed search or protection against benchmark indexing, we believe other web-search based agents should face similar risks.

\section*{Ethics Considerations}
This work could introduce dual-use risks, as the proposed methodology could be misused to manipulate evaluations of search-based agents by deliberately creating search-time contamination through online benchmark leakage. We acknowledge this concern, but we believe that transparent disclosure is necessary for revealing systematic weaknesses in current evaluation protocols and improving the integrity of agent benchmarks. Our goal is to support contamination-aware evaluation rather than to facilitate benchmark exploitation or artificially improve agent rankings.

\bibliography{main}

\appendix

\section{Time-varying Cox Proportional Hazards Model}
\label{apd:cox}

To jointly estimate the effects of different contamination types, we employ a multivariable Cox proportional hazards model with time-dependent covariates:
\begin{equation}
\resizebox{0.95\linewidth}{!}{$
h(t|\mathbf{x}(t)) = h_0(t) \exp\left(
\beta_{1} x_{1}(t) +
\beta_{2} x_{2}(t) +
\beta_{3} x_{3}(t)
\right)$},
\end{equation}
where:
\begin{itemize}[topsep=1pt,leftmargin=15pt,itemsep=0pt]
\item $t$ denotes the turn within the trajectory.
\item $h(t|\mathbf{x}(t))$ is the hazard rate, representing the conditional probability that the agent reaches a correct prediction at turn $t$, given that it has not reached a correct prediction before turn $t$.
\item $h_0(t)$ is the baseline hazard of reaching a correct prediction at turn $t$ when no STC covariates are active.
\item $x_{1}(t)$, $x_{2}(t)$, and $x_{3}(t)$ are binary time-dependent covariates indicating whether BML, QCL, or EAL has occurred by turn $t$.
\item $\beta_{1}$, $\beta_{2}$, and $\beta_{3}$ are regression coefficients. The hazard ratio for each STC type is given by $\exp(\beta_k)$, which quantifies its independent association with the likelihood of reaching a correct prediction, controlling for the other STC types.
\end{itemize}

\textit{For example, a remarkably high HR for \stypethree contamination indicates that the agent outputs the correct prediction quickly after answers are explicitly leaked. Conversely, a lower HR implies a delayed or weaker correlation between the contamination event and the successful prediction.}

\section{Dataset Description}
\label{sec:dataset}

\begin{itemize}[topsep=1pt,leftmargin=5pt,itemsep=0pt]

\item \textbf{PubMedQA} \cite{jin-etal-2019-pubmedqa} is a biomedical question answering dataset collected from PubMed abstracts. The task involves answering research questions with yes/no/maybe using the corresponding article abstracts as context. We evaluate on the PQA-L test set, consisting of 500 questions. It requires interpreting complex quantitative information and comparing groups. 

\item \textbf{MedQA}~\cite{jin-2021-medqa} is a multilingual medical question answering dataset compiled from medical board exam questions. Following prior works~\citep{singhal-2023-llmclinical, nori2023can}, we evaluate on the English subset corresponding to the United States Medical Licensing Examination (USMLE), using 1273 test questions in a four-option multiple-choice format. These questions require clinical knowledge, deep understanding of medical concepts and multi-hop reasoning. 

\item \textbf{MMLU-Med} is a medical subset of the Massive Multitask Language Understanding (MMLU) benchmark~\cite{hendrycks2021measuring}, which contains multiple-choice questions spanning 57 subjects across humanities, sciences, and professional domains. Following prior work~\citep{singhal-2023-llmclinical}, we evaluate the test sets of six medical subjects (\textit{College Medicine, College Biology, Professional Medicine, Anatomy, Medical Genetics, and Clinical Knowledge}), comprising a total of 1089 four-option questions. we exclude the shortest half of questions, as they are more likely to require only factual recall rather than complex reasoning. 

\item \textbf{MedMCQA}~\cite{pal-2022-medmcqa} is a large-scale medical QA dataset sourced from Indian postgraduate medical entrance exams (AIIMS PG and NEET PG). The evaluation set comprises 4183 four-option multiple-choice questions. Same with MMLU, we also exclude the shortest half of questions which only require the simple factual recall. 

\textbf{MedXpertQA} \cite{zuo2025medxpertqa} is a medical 10-choice question answering benchmark collected from professional medical licensing and specialty board exams. We evaluate on the MedXpertQA Text test set, consisting of 2,450 questions.  The 10-option settings forces models to demonstrate genuine clinical knowledge rather than relying on statistical luck, making the evaluation metric far more rigorous and reliable.

\textbf{HLE-149} \cite{phan2025humanity} is an expert-level biological and chemical QA dataset curated from the Humanity's Last Exam (HLE) benchmark. It was constructed by strictly auditing the original HLE subset to remove erroneous, ambiguous, or literature-conflicting items. In our experiments, we evaluate on this highly verified gold split, which comprises 149 scientifically rigorous questions.

\textbf{MedBullets}~\cite{chen-2025-benchmarking} is a multiple-choice medical question answering dataset consisting of 308 USMLE Step 2\&3 style practice questions, which were originally posted by MedBullets on Twitter/X between April 2022 and December 2023. Each question features a detailed patient case vignette accompanied by five answer choices, the correct answer key, and a professionally written explanation by medical experts. Different from traditional benchmarks that focus on textbook knowledge recall, MedBullets is specifically designed to evaluate complex, multi-step clinical reasoning capabilities of models.
\end{itemize}

As shown in Table \ref{tab:datasets}, these datasets cover a diverse range of medical QA formats, including multiple-choice questions, and exact-match answers. In total, our evaluation includes 6,803 samples from benchmarks released between 2020 and 2025, allowing us to examine search-time contamination across both established and recently released medical evaluation sets. 

\begin{table}[tb]
\centering
\caption{Summary statistics of the evaluated medical benchmarks. 
MCQ = multiple-choice question; EM = exact match.}
\resizebox{0.48\textwidth}{!}{
\label{tab:datasets}
\begin{tabular}{lccc}
\toprule
Name &  \#Samples & Question Type & Year \\
\midrule
MedQA & 1273 & MCQ & 2020\\
MMLU & 544 & MCQ & 2021\\
MedMCQA & 2089 & MCQ  & 2022\\
MedXpertQA & 2450 & MCQ & 2025\\
HLE-149 & 149 & MCQ \& EM &2025 \\
Medbullets5op & 298 & MCQ & 2025 \\
\bottomrule
\end{tabular}}
\end{table}

\section{Deep Research Agents Description}
\label{apd:agents}
\begin{table*}[htbp]
\centering
\caption{Comparison of evaluated deep research agents.}
\label{tab:agent_comparison}
\resizebox{\linewidth}{!}{
\begin{tabular}{l|cccc}
\toprule
\textbf{Agent} & \textbf{Agent Model} & \textbf{Search Engine} & \textbf{Web Browser} & \textbf{Search Language} \\
\midrule
Tongyi Deep Research 
& Tongyi-DeepResearch-30B-A3B & Serper & Jina & English \\

Step Deep Research 
& Step 3.5 Flash 
& Built-in Web search 
& Built-in Web Reader
& Multilingual \\

Gemini Deep Research 
& Gemini 3.1 Pro
& Google Search
& Built-in Web Reader
& Multilingual \\

Valyu Deep Research 
& Valyu API 
& Valyu AI-native search 
& Built-in Web Reader 
& NA \\
\bottomrule
\end{tabular}}
\end{table*}
We evaluate a representative set of deep research agents spanning both open-source models and commercial research systems: 
\begin{itemize}[topsep=1pt,leftmargin=5pt,itemsep=0pt]
\item \textbf{Qwen3-30B-A3B }\cite{qwen3technicalreport}: the base model on which Tongyi Deep Research is fine-tuned. We include it to assess the model capability before applying the full deep research agent pipeline.
\item \textbf{Tongyi Deep Research} \cite{team2025tongyi}: a publicly available deep research agent that provides detailed execution traces, including search queries, retrieved webpages, intermediate summaries, and reasoning details. 
\item \textbf{Step Deep Research} \cite{hu2025stepdeepresearchtechnicalreport}: a cost-effective commercial agent for autonomous information gathering and professional report generation in open-ended research scenarios. 
\item \textbf{Gemini Deep Research} \cite{google2026gemini}: a closed-source commercial deep research system that provides only the visited URLs and summarized retrieved content, with limited visibility into its web search process. 
\item \textbf{Valyu Deep Research} \cite{valyu2026deepresearch}: an API-based deep research agent built on Valyu's AI search infrastructure, which is reported to achieve leading performance on web search benchmarks. We evaluate its fast modes to prioritize cost-efficiency and low latency. 
\end{itemize}

The comparison of evaluated deep research agents are listed in Table \ref{tab:agent_comparison}. Our main experiments are conducted on Tongyi Deep Research because it is cost-effective, publicly accessible, and provides fully observable execution traces, allowing us to inspect search queries, visited webpages, intermediate summaries, and reasoning steps. To examine the generalizability of our findings, we further evaluate several other deep research agents. 

\section{Human-Automatic Judge Agreement}
\label{apd:human_automatic}

\begin{table}[tb]
\centering
\caption{Turn-level human evaluation results of \stypethree detection.}
\resizebox{0.35\textwidth}{!}{
\label{tab:human_eval}
\begin{tabular}{lcc}
\toprule
Dataset & Precision & Recall \\
\midrule
MedQA & 94.87\% & - \\
Medbullets5op & 100\% & 83.33\% \\
\bottomrule
\end{tabular}}
\end{table}

To validate the accuracy of automatic judgments, we conduct manual annotation of the \stypethree detection results on two representative datasets, as shown in Table \ref{tab:human_eval}. For MedQA, we manually review the automatically detected \stypethree cases and assess whether they satisfied the predefined criteria. For MedBullets5op, we manually annotate the first 160 questions and use these annotations as reference labels to estimate both the precision and recall of the automatic detection results on this subset. Overall, our detection algorithm shows substantial agreement with human annotations, suggesting its reliability for automated \stypethree identification.

\section{Experiment Costs}
Table \ref{tab:cost_summary} summarizes the API costs of our experiments. 
The total cost is mainly composed of two parts: agent execution and LLM-as-a-judge evaluation.  Due to these costs, we restrict commercial-agent evaluation to two datasets with 200 questions in total. 
Scaling the same protocol to all benchmarks and all commercial agents would be prohibitively expensive. 
Even so, commercial deep research systems are costly: Gemini DeepResearch alone costs 200 USD, while Valyu Fast and Step DeepResearch cost 35 USD and 18 USD, respectively. 

Therefore, our full-scale analysis is conducted mainly on Tongyi DeepResearch, which provides observable search trajectories and a more controllable evaluation pipeline, while commercial systems are used to assess whether the observed STC patterns generalize beyond a single agent.

\begin{table}[tb]
\centering
\caption{Experiment Cost.}
\label{tab:cost_summary}
\resizebox{0.4\textwidth}{!}{
\begin{tabular}{lc}
\toprule
Agent / Component & Cost (USD) \\
\midrule
Tongyi DeepResearch & 384 \\
Gemini DeepResearch & 200 \\
Valyu [Fast] & 35 \\
Step DeepResearch & 18 \\
LLM Judge: DeepSeek v4 Pro & 221 \\
\midrule
\textbf{Total} & \textbf{858} \\
\bottomrule
\end{tabular}}
\end{table}

\begin{figure*}
    \centering
    \begin{tcolorbox}[
      colback=white, colframe=black, arc=1mm, width=1\linewidth,
      title=\textbf{Prompt for Detecting Explicit Answer Leakage}, 
      coltitle=white, colbacktitle=orange, fonttitle=\bfseries
    ]
You are a strictly objective Data Contamination Auditor. \\

Your task is to analyze `Web Search Content' to determine if an AI system bypassed genuine reasoning through ``Search-Time Contamination''—retrieving the exact benchmark question and its corresponding answer key directly from the web. \\

Inputs you will receive: \\
1. [Question]: The original benchmark question.

2. [Ground-Truth Answer]: The correct answer.

3. [Web Search Content]: The retrieved context to be evaluated. \\

Your goal is to distinguish between ``Search-Time Contamination'' (contaminated=true) and ``Legitimate Knowledge Retrieval'' (contaminated=false). \\

\#\#\# Evaluation Criteria: \\
Set `contaminated = true' ONLY IF BOTH of the following strict conditions are simultaneously met: \\
1. EXACT Question Fingerprint Match: The search content contains extended, contiguous blocks of verbatim text from the original question. \\
- EXCLUSION (Crucial): It MUST NOT be a merely ``similar'' or analogous question. If the content describes a parallel scenario but uses different variables, numbers, patient demographics (e.g., age, gender), or specific lab values, it is NOT an exact match. Mere keyword overlap is insufficient. \\

2. DIRECT Answer Leakage: The exact Ground-Truth Answer or correct option label MUST be explicitly paired with the matched question text in the search content (e.g., within the same Q\&A block, flashcard format, or forum post). \\
- EXCLUSION (Crucial): If the web content contains the exact question but DOES NOT explicitly provide the ground-truth answer key, you MUST set `contaminated = false'. Retrieving an unanswered question does not constitute answer leakage. \\

Set `contaminated = false' (Legitimate Knowledge Retrieval) in ALL other cases, strictly including when: \\
- The content is a similar practice question with altered details or different variables. \\
- The content contains the exact question but lacks the answer.\\
- The content provides general encyclopedic knowledge, literature snippets, or clinical guidelines that help *deduce* the answer, but lacks the exact Q\&A pair.\\

\#\#\# Output Format: \\
You must output a strictly valid JSON object with the following structure: \\
\{ \\
\hspace*{2em}  ``reasoning'': ``Step 1: Check for EXACT string matching (reject similar questions with different variables). Step 2: Check if the ground-truth answer is explicitly provided alongside the question. Step 3: Conclude contamination status based on the strict AND logic.'', \\
\hspace*{2em} ``contaminated'': true or false \\
\}
\end{tcolorbox}
\caption{Prompt for Detecting Explicit Answer Leakage.}
    \label{fig:prompt_type3}
\end{figure*}

\section{PubMedQA Leakage Rate on ValYu}
\label{apd:valyu}
Valyu presents a distinctive case among the evaluated agents. Unlike general-purpose web search engines, Valyu employs a proprietary AI-native search infrastructure that integrates curated data sources, including professional knowledge repositories. In our MedQA evaluation, we observed no STC for Valyu, suggesting that its constrained retrieval backend mitigates exposure to widely redistributed benchmark artifacts. 

However, given that Valyu has supported PubMed retrieval, we hypothesized that STC might resurface if the benchmark itself originates from PubMed. To investigate this, we evaluated Valyu on 100 samples from PubMedQA. Strikingly, 65 of these questions exhibited STC. This finding demonstrates that STC vulnerability is not merely a dichotomy between open-web and restricted-source systems. Rather, the risk is fundamentally driven by the overlap between the agent's retrieval corpus and the benchmark's source material; when a benchmark is derived from a searchable corpus, even highly specialized search infrastructures remain susceptible to severe contamination. 

\section{URLs in \stypeone Detection}
\label{sec:urls}

To detect \typeone metadata leakage, we employ URL-level matching as a lightweight and interpretable signal, shown in Table \ref{tab:url-matching-patterns}. We categorize sensitive URL patterns into two distinct groups. The first group targets common hosts of benchmark artifacts, encompassing open-data platforms (e.g., Hugging Face and GitHub) alongside exam-preparation, flashcard, document-sharing, and Q\&A websites (e.g., CourseHero, Quizlet, Scribd, MedicoApps, and MedBullets). These platforms frequently distribute redistributed benchmark instances, exam-style queries, and answer discussions. The second group comprises benchmark-specific identifiers, including dataset names, paper URLs, and known source patterns such as \texttt{MedQA}, \texttt{MedMCQA}, \texttt{MMLU}, \texttt{MedXpertQA}, \texttt{HLE}, and \texttt{MedBullets}. If a retrieved URL matches any of these patterns, we flag it as potential metadata leakage, indicating that the agent's search trajectory has been misdirected toward benchmark artifacts rather than authentic, task-relevant clinical knowledge sources.

\begin{table*}[h]
\centering
\caption{URL matching patterns used for sensitive website detection.}
\label{tab:url-matching-patterns}
\resizebox{\linewidth}{!}{
\begin{tabular}{lll}
\toprule
Type & Matching keywords & Description \\
\midrule
\multicolumn{3}{c}{Common Matching Keywords} \\
\midrule
\multirow{2}{*}{Data-hosting Platform}
& {\scriptsize\texttt{huggingface.co/datasets}} & Hugging Face open-source datasets. \\
& {\scriptsize\texttt{github.com}} & GitHub open-source repositories. \\
\midrule
\multirow{23}{*}{Medical Exam URL}
& {\scriptsize\texttt{coursehero.com}} & Study resources and Q\&A pages. \\
& {\scriptsize\texttt{getoncourse.ai}} & AI-based medical exam preparation website. \\
& {\scriptsize\texttt{medicoapps.org}} & Indian medical exam preparation website. \\
& {\scriptsize\texttt{accessmedicine.mhmedical.com/...question...}} & Medical review questions or quizzes. \\
& {\scriptsize\texttt{quizlet.com}} & Flashcards, quizzes, and practice tests. \\
& {\scriptsize\texttt{cram.com/flashcards}} & Flashcard-based study materials. \\
& {\scriptsize\texttt{scribd.com/document}} & Uploaded study or exam documents. \\
& {\scriptsize\texttt{orthobullets.com/testview}} & Orthopedic exam-style question pages. \\
& {\scriptsize\texttt{dentaldevotee.com/...yyyy...}} & Dental MCQs and past questions. \\
& {\scriptsize\texttt{transtutors.com/questions}} & Solved homework Q\&A. \\
& {\scriptsize\texttt{cliffsnotes.com/cliffs-questions}} & Homework Q\&A solutions. \\
& {\scriptsize\texttt{homework.study.com/explanation}} & Expert step-by-step explanations. \\
& {\scriptsize\texttt{AIIMS}} & Indian medical entrance-exam content. \\
& {\scriptsize\texttt{dental pulse}} & Dental exam preparation materials. \\
& {\scriptsize\texttt{SMART DENTAL REVISION}} & Dental revision resources. \\
& {\scriptsize\texttt{ORAL SURGERY LIVE Test}} & Oral surgery test materials. \\
& {\scriptsize\texttt{USMLE}} & Medical licensing exam content. \\
& {\scriptsize\texttt{osmosis.org/blog/usmle}} & Medical licensing exam study resources. \\
& {\scriptsize\texttt{passmed}} & Medical exam question banks. \\
& {\scriptsize\texttt{MCQ}} & Multiple-choice questions. \\
& {\scriptsize\texttt{quiz}} & Quiz-based materials. \\
& {\scriptsize\texttt{question}} & Question-based content. \\
& {\scriptsize\texttt{test}} & Test-oriented content. \\

\midrule
\multicolumn{3}{c}{Dataset-adapted Matching Keywords} \\
\midrule
\multirow{4}{*}{MedBullets5op}
& {\scriptsize\texttt{step2.medbullets.com/testview}} & TestView question pages. \\
& {\scriptsize\texttt{facebook.com/medbullets}} & Related Facebook pages. \\
& {\scriptsize\texttt{medbullets}} & Keyword-based URL matches. \\
& {\scriptsize\texttt{medbullets\_op4 / medbullets\_op5}} & Dataset identifier matches. \\
\midrule
\multirow{2}{*}{MedQA}
& {\scriptsize\texttt{medqa}} & Keyword-based URL matches. \\
& {\scriptsize\texttt{OpenMedQA}} & Open-version keyword matches. \\
\midrule
\multirow{3}{*}{MedMCQA}
& {\scriptsize\texttt{dokumen.pub}} & Document-hosting pages. \\
& {\scriptsize\texttt{facebook.com}} & Related Facebook pages. \\
& {\scriptsize\texttt{medmcqa}} & Keyword-based URL matches. \\
\midrule
MMLU
& {\scriptsize\texttt{mmlu}} & Keyword-based URL matches. \\
\midrule
\multirow{2}{*}{MedXpertQA}
& {\scriptsize\texttt{arxiv.org/pdf/2505.18283}} & Article containing leaked questions. \\
& {\scriptsize\texttt{MedXpertQA}} & Keyword-based URL matches. \\
\midrule
\multirow{2}{*}{HLE-149}
& {\scriptsize\texttt{HLE}} & Keyword-based URL matches. \\
& {\scriptsize\texttt{Humanity's Last Exam}} & Full-name keyword matches. \\
\bottomrule
\end{tabular}
}

\end{table*}

\end{document}